\documentclass[apjl]{emulateapj}
\usepackage{apjfonts}

\usepackage{color}

\shorttitle{Young population in the Monoceros ring}
\shortauthors{Carballo-Bello et al.}

\newcommand{\figref}{Figure~\ref}

\newcommand{\secref}{Section~\ref}

\begin{document}

\title{A Megacam Survey of Outer Halo Satellites. IV. Two foreground populations possibly associated with the Monoceros Substructure in the direction of NGC\,2419 and Koposov\,2\altaffilmark{1}}
\author{Julio A. Carballo-Bello\altaffilmark{2},  
Ricardo R. Mu\~noz \altaffilmark{2}, 
Jeffrey\ L.\ Carlin \altaffilmark{3}, 
Patrick\ C\^ot\'e\altaffilmark{4}, 
Marla\ Geha\altaffilmark{5}, 
Joshua\ D.\ Simon\altaffilmark{6}, 
S.\ G.\ Djorgovski\altaffilmark{7}
}

\altaffiltext{1}{Based on observations obtained at the Canada-France-Hawaii 
Telescope (CFHT) which is operated by the National Research Council of 
Canada, the Institut National des Sciences de l'Univers of the Centre 
National de la Recherche Scientifique of France,  and the University of Hawaii.}

\altaffiltext{2}{Departamento de Astronom\'ia, Universidad de Chile, 
Camino del Observatorio 1515, Las Condes, Santiago, Chile (jcarball@das.uchile.cl)}

\altaffiltext{3}{Department of Physics, Applied Physics and Astronomy, Rensselaer Polytechnic Institute, Troy, NY 12180, USA}

\altaffiltext{4}{National Research Council of Canada, Herzberg Astronomy and Astrophysics, Victoria, BC, V9E 2E7, Canada}

\altaffiltext{5}{Astronomy Department, Yale University, New Haven, CT 06520, USA}

\altaffiltext{6}{Observatories of the Carnegie Institution of Washington, 
813 Santa Barbara St., Pasadena, CA 91101, USA}

\altaffiltext{7}{Astronomy Department, California Institute of Technology, 
Pasadena, CA, 91125, USA}

\begin{abstract}

The origin of the Galactic halo stellar structure known as the Monoceros ring is still under debate. 
In this work, we study that halo substructure using deep CFHT wide-field photometry obtained for the globular clusters 
NGC\,2419 and Koposov\,2, where the presence of Monoceros becomes significant because of their coincident projected position. Using Sloan Digital Sky Survey photometry and spectroscopy in the area surrounding these globulars and beyond, where the same Monoceros population is detected, we conclude that a second feature, not likely to be associated with  Milky Way disk stars along the line-of-sight, is present as foreground population. Our analysis suggests that the Monoceros ring might be composed of an old stellar population of age $t \sim 9$\,Gyr and a new component $\sim 4$\,Gyr younger at the same heliocentric distance. Alternatively, this detection might be associated with a second wrap of Monoceros in that direction of the sky and also indicate a metallicity spread in the ring. The detection of such a low-density feature in other sections of this halo substructure will shed light on its nature.

\end{abstract}
\keywords{Galaxy: structure, halo, globular clusters: general }

\section{Introduction}

Among the stellar substructures discovered so far in the Milky Way halo, the so-called Monoceros ring is one of the most challenging structures for Galactic archeology. Unveiled by 
\cite{Newberg2002} and \cite{Yanny2003} in Sloan Digital Sky Survey \citep[SDSS,][]{York2000} data as a stellar overdensity at low Galactic latitudes, its nature is still controversial, despite substantial observational efforts \citep[e.g.][]{Conn2007,Conn2008,Slater2014}.
One of the leading interpretations is that the Monoceros ring is the remnant of a past accretion event \citep[e.g.,][]{Conn2005,Juric2008,Chou2010,Sollima2011}, similar to that generated by the disruption of the Sagittarius (Sgr) dwarf galaxy, which is orbiting around our Galaxy in an almost polar orbit \citep[e.g.][]{Ibata1994,Majewski2003,Bonifacio2004,MartinezDelgado2004,Bellazzini2006a,Belokurov2006,Siegel2007,Koposov2012}. In contrast to Sgr, Monoceros lacks a known progenitor system, although it has been proposed and later discarded, that the Canis Major overdensity is the accreted system that formed Monoceros \citep{Martin2004,Momany2004,Martinez-Delgado2005,Momany2006,Moitinho2006,Bellazzini2006a,Butler2007,Mateu2009}. 

An alternative scenario presents the Monoceros ring as the result of a distortion of the Galactic plane \citep{Momany2004,Momany2006,Hammersley2011}. These studies suggest that the observed star counts are reproducible considering a flared thick disk without a cut-off at $R \sim 14$\,kpc. However, \cite{Sollima2011} has recently shown that none of the available synthetic models for the Milky Way are able to reproduce the observed stellar counts in the Monoceros ring. Unfortunately, none of the existing arguments favoring or rejecting the extragalactic origin of the Monoceros ring rule out completely the other hypotheses. Additional processes have been suggested to explain the detection of such a vast halo substructure, including the disk distortion generated by a close encounter \citep{Younger2008}, the existence of caustic rings of dark matter in that position within the Galaxy \citep{Natarajan2007} and the accretion of the Sgr dwarf galaxy, that might have a direct impact in the formation of stellar rings in the outer halo \citep{Michel-Dansac2011,Purcell2011}.

Different spectroscopic studies have reported metallicities for Monoceros in the range $-1.6 <$ [Fe/H] $< -0.4$ \citep[e.g.][]{Crane2003,Rocha-Pinto2003,Yanny2003} but recent estimates converged on [Fe/H] $\sim -1.0$ with a relatively low dispersion \citep{Ivezic2008,Conn2012,Meisner2012}. As for the age of the ring, \cite{Sollima2011} derived a value of  $t = 9.2 \pm 0.2$\,Gyr via isochrone fitting in two fields in the anticenter direction. 

In this work, we have used  deep and wide-field photometry for the globular clusters (GCs) NGC\,2419 and Koposov\,2 (Kop\,2) obtained in the context of a larger photometric survey, to study the stellar populations associated with the Monoceros ring. NGC\,2419 and Kop\,2 are located at $d_{\odot} = 83.2$ and $\sim40$\,kpc respectively \citep{Koposov2007,Ripepi2007} in the anticenter region, where an important amount of stellar structures (potentially different to Monoceros) have been found \citep{Grillmair2006d,Grillmair2008,Li2012}. The projected positions of the clusters in the sky are consistent with the orbit for the Monoceros ring proposed by the \cite{Penarrubia2005} model.

\begin{figure}[th]
     \begin{center}
      \includegraphics[scale=1.5]{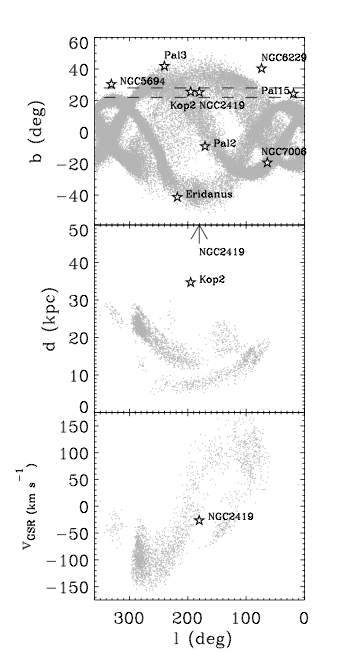}
      \caption[CMDs]{\small \emph{Upper panel}: tidal debris spatial distribution of the Monoceros ring as proposed by the \cite{Penarrubia2005} model. The position of the GCs included in the Megacam survey and with $|b| < 45^{\circ}$ are indicated as stars \citep{Harris2010}. \emph{Middle and bottom panels}: predicted heliocentric distance and velocities (Galactic standard of rest) for the $23 < b < 28^{\circ}$ section of the ring as defined by the dashed lines in the upper panel, respectively. There is not measured radial velocity for Kop\,2.} 
\label{monoceros_model}
     \end{center}
    \end{figure}

\section{Observations}

The data used for this study are part of a larger photometric survey of all outer Galactic halo satellites (R. R. Mu\~noz et al., {\it in preparation}) carried out with the Megacam imagers at both the Canada-France-Hawaii Telescope (CFHT)
in the north and the Magellan II-Clay Telescope in the south. In the particular case of NGC~2419 and Kop~2, the observations were made
at CFHT.

\begin{figure*}[th]
     \begin{center}
      \includegraphics[scale=1.5]{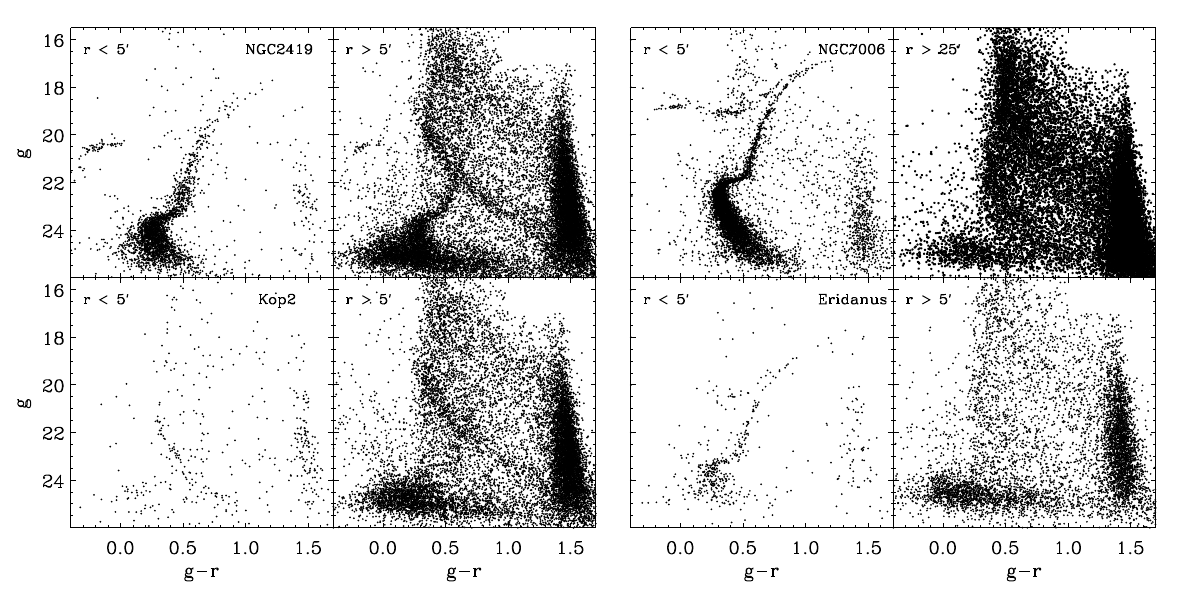}
      \caption[CMDs]{\small CMDs corresponding to the inner 5\,arcmin stellar content and those stars beyond that distance from the centers of NGC\,2419 and Kop\,2 (left panels), and NGC\,7006 and Eridanus (right panels). Note that in the case of NGC\,7006, we have selected those stars beyond 25\,arcmin to reduce the number of field stars. The $r > 5$\,arcmin CMDs in the cases of NGC\,2419 and Kop\,2 show the presence of a  narrow subjacent MS in the range $19 < g < 24$, possibly associated with the Monoceros ring. That feature is not detected in the NGC\,7006 and Eridanus diagrams.}
\label{cmdscomparative}
     \end{center}
    \end{figure*}

Observations with the MegaCam imager on the CFHT were made in queue mode. MegaCam is a wide-field imager consisting of $36$ $2048\times4612$ pixel CCDs, covering almost a full $1\times1$ deg$^{2}$ field of view with a pixel scale of $0\arcsec.187$/pixel.
For each object shown in Figure~\ref{cmdscomparative} one pointing was observed.  For each pointing, six dithered exposures in 
SDSS $g$ and $r$ in mostly dark conditions were observed, with typical seeing of $0\arcsec.7-0\arcsec.9$. The individual exposure times ranged between 240 and 500\,s in both $g$ and $r$. Table~\ref{t:list} lists a summary of the observing logs for these four objects,
including their center coordinates, average air masses in the $g$ and $r$ filters. The dithering pattern was selected from the standard MegaCam operation options in order to cover both the small and large gaps between chips (the largest vertical gaps in MegaCam 
are six times wider than the small gaps). 
Point source photometry was carried out using both DAOPHOT/Allstar and ALLFRAME \citep{Stetson1994} as detailed in \citet{Munoz2010}.
The astrometric solutions present in the headers of the images were refined using the freely available
SCAMP\footnote{See http://astromatic.net/software/scamp/} package. Photometric calibration was performed by directly comparing stars 
in SDSS (DR7) for those objects in the SDSS footprint, and using SDSS fields as secondary standard calibrators for those objects outside.\\

  \begin{figure*}[th]
     \begin{center}
      \includegraphics[scale=1]{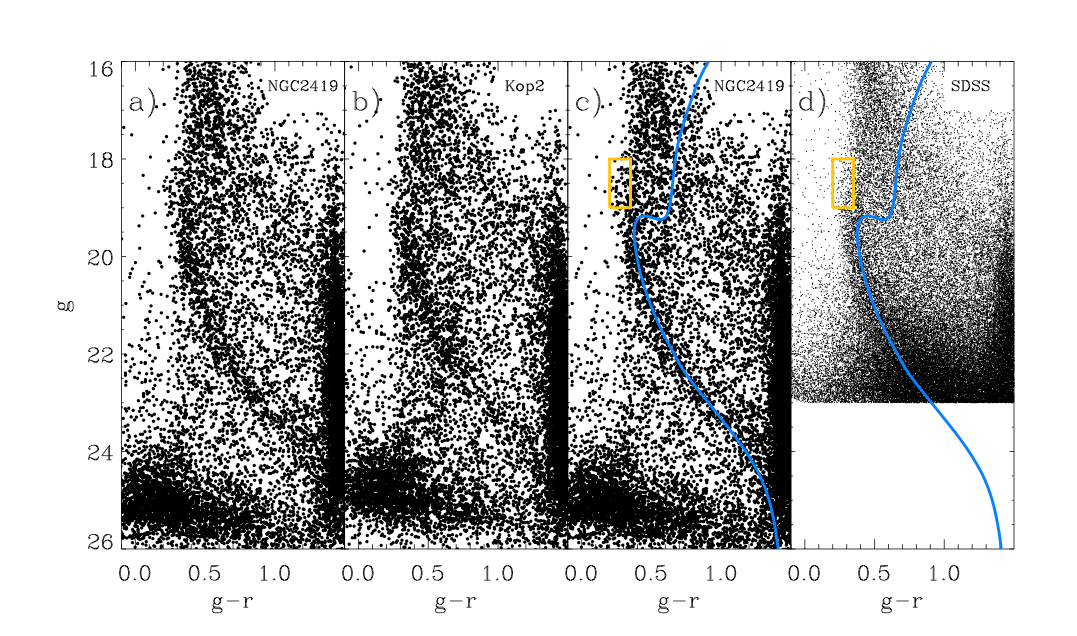}
      \caption[CMDs]{{\small {\it Panels (a) and (b):} CMDs corresponding to those 
      stars beyond $r= 10$ and $3$\,arcmin from the centers of NGC\,2419 and Kop\,2 respectively. {\it Panel (c):} CMD corresponding to the surroundings of NGC\,2419, where the Monoceros isochrone has been overplotted in blue and the position of the BP analyzed in this work is indicated by a yellow box. {\it Panel (d):} CMD obtained as the sum of the objects in 5 SDSS fields between NGC\,2419 and Kop\,2. Same isochrone and BP position are included for comparison.}} 
\label{cmds}
     \end{center}
   \end{figure*}

\section{Results and discussion}

\subsection{Color-magnitude diagrams}

Around half of the original targets of our primary survey were GCs. Interestingly, NGC\,2419, Kop\,2, NGC\,7006 and Eridanus lie in the predicted direction of the halo substructure known as the 
Monoceros ring according to the \cite{Penarrubia2005} model (\figref{monoceros_model}) and whose presence should be revealed by a foreground main-sequence (MS) in the respective color-magnitude diagrams (CMDs). \figref{cmdscomparative} shows the CMDs containing both stars within a $5'$ radius area and those further away that distance from the center of those clusters. Clear MSs are revealed in the CMDs for the surrounding areas of NGC\,2419 and Kop\,2 in the ranges $19 < g < 24$ and $ 0.3 < g-r < 1.3$ (right panels). As it is clear from the CMDs, the morphology of these MSs differ from the observed features associated with the GC populations (left panels). These diagrams confirm that NGC\,2419 and Kop\,2 lie in the same light-of-sight of the vast stellar halo substructure which we identify as the Monoceros ring. In the case of Eridanus, it is not possible to identify any features possibly associated with a population different to those associated with the Milky Way components. Eridanus might be still surrounded by a low surface-brightness region of Monoceros but a complete absence of such substructure is also a possibility. As for NGC\,7006, the CMD suggests the presence of an overdensity in its $g > 21$ section. That subjacent MS might be associated, as proposed by \cite{Carballo-Bello2014}, to the presence of stars belonging to the Hercules-Aquila cloud \citep{Belokurov2007a,Simion2014} along the line-of-sight to the cluster. Alternatively, a higher density of halo stars in that direction of the sky as suggested by \cite{Deason2014} results might produce a broader halo MS as the one observed in our data. In this work we will focus on the well-defined foreground populations around NGC\,2419 and Kop\,2 to explore Monoceros.

Since this article does not focus on the clusters themselves, we minimize the number  
of GC stars in the resulting diagrams, and therefore in making the CMDs to detect and study 
Monoceros we have included only those stars away from the cluster's centers.
The King tidal radii of NGC\,2419 and Kop\,2 are $r_{\rm t}= 7'.5$ and $0'.8$  respectively 
(R. R. Mu\~noz et al., {\it in preparation}), although this parameter does not  indicate necessarily 
the region beyond which cluster stars are no longer present 
\citep[see discussion in][]{Carballo-Bello2012}. Thus, to
work with the cleanest possible CMDs we select stars beyond $r=10'$ and $3'$ from 
NGC\,2419 and Kop\,2's centers respectively. 

The {\it a)} and {\it b)} panels of \figref{cmds} show the resulting CMDs for the area surrounding NGC\,2419 and Kop\,2, respectively. 
 The observed MS widths indicate the detection of a stellar structure in a narrow distance range along the line-of-sight. We note that the foreground sequences seem to extend  blueward of the disk stars turnoff (TO) at $0.20 < g-r < 0.35$ and $18 < g < 19$, with a much larger separation with respect to the tentative Monoceros TO than the photometric error at this level ($|\delta g| \sim 0.02$). This feature is observed in both fields, although it is slightly clearer in the NGC\,2419 data. To investigate whether this blue population (herein BP) corresponds to the main Monoceros population, we have visually fitted a theoretical 
isochrone \citep{Dotter2008} corresponding to the nominal Monoceros age, determined to be $\sim 9$\,Gyr by \cite{Sollima2011} and with a metallicity of [Fe/H]$=-1$ \citep{Ivezic2008,Conn2012,Meisner2012}. The adopted {\it{E(B-V)}} values in the direction of NGC\,2419 and Kop\,2  are 0.035 and 0.037\,mag respectively \citep {Schlafly2011}. 
As shown in {\it c)} panel of \figref{cmds}, this isochrone reproduces the morphology of the foreground MS reasonably well but fails
to cover them all the way up to the blue end.
By using the region of the isochrone that matches the observed MSs
we obtained a radial distance of $d_{\odot}=10.2\pm1.6$ and $10.0\pm1.5$\,kpc for the 
underlying system in the surroundings of NGC\,2419 and Kop\,2, respectively. These estimates are consistent with the predictions made by the \cite{Penarrubia2005} model for the Monoceros ring in that line-of-sight and with the heliocentric distance derived by \cite{Li2012} for nearby fields on that structure. From the projected position of the clusters on the Fig.\,14 on \cite{Li2012}, it is possible to rule out the so-called Anticenter stream discovered by \cite{Grillmair2006d} as the subjacent population in the fore/background of these clusters.

\subsection{SDSS analysis}

  \begin{figure}[h!]
     \begin{center}
      \includegraphics[scale=1.35]{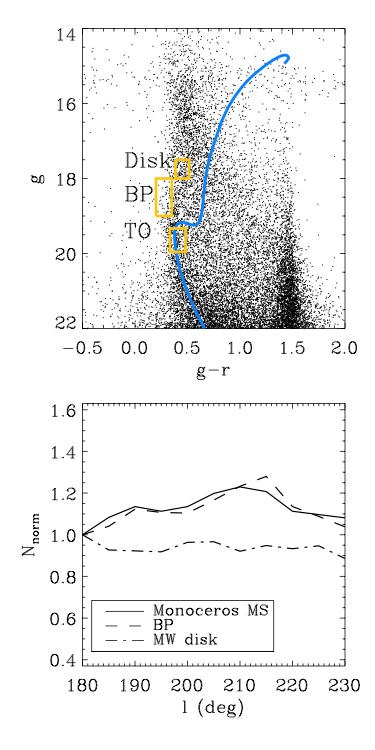}
      \caption[CMDs]{{\small {\it Upper panel:} the 
      diagram corresponding to the SDSS field centered at $(l,b) = (184, 25.35)^{\circ}$. 
      The overplotted blue isochrone corresponds to the Monoceros ring. {\it Bottom panel:} variation of the stellar counts for the 3 regions defined in the upper CMD (disk, BP and TO sample stars indicated as yellow rectangles) as a function of the Galactic longitude in the range $180 < \ell < 230^{\circ}$ at $b=25^{\circ}$. 
      All the sequences are normalized to the first datapoint.}}
\label{comparative}
     \end{center}
    \end{figure}

  \begin{figure}[t]
     \begin{center}
      \includegraphics[scale=0.65]{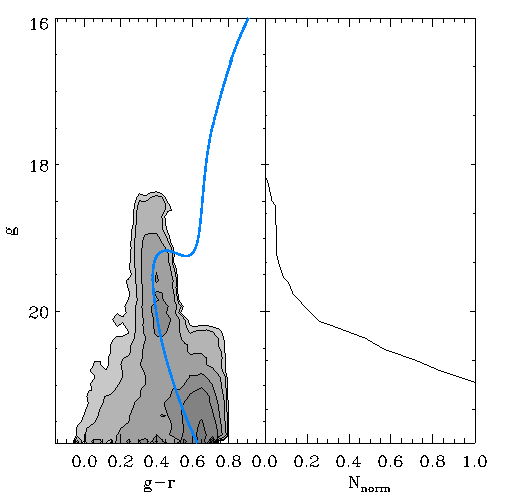}
      \caption[CMDs]{{\small  \emph{Left:} Hess diagram corresponding to the CMD obtained at $(l,b)=(210,25)^{\circ}$, when the Galactic components are removed by the substraction of the CMD corresponding to a field centered at $(l,b)=(180,25)^{\circ}$. The blue isochrone represents the Monoceros ring stellar population and the minimal density contour indicates the bins with a $4\%$ of the highest density obtained. \emph{Right:} cumulative luminosity function derived from the Hess diagram and for the color range $0.3<g-r<0.5$. The values have been normalized by the last datapoint considered at $g = 22$.}}
\label{hess}
     \end{center}
    \end{figure}

To assess the possibility that these blue extensions of the Monoceros MSs 
do not represent an actual Monoceros population but instead correspond
to statistical fluctuations of Milky Way stars in this part of the CMD,  we have explored 
an expanded area around the clusters using SDSS data. 
The right panel in \figref{cmds} shows all the objects within 
a radius of $60'$ in five fields in the coordinate range $(\ell, b)=(182,25.35)^{\circ}$ 
to $(190,25.35)^{\circ}$, which are located between NGC\,2419 and Kop\,2. 
Though SDSS photometry is shallower than our CFHT data, the Monoceros MS is clearly visible.
 The blue extension observed in the CFHT data is also clear in this CMD.

To establish whether the origin of the BP is Monoceros or Milky Way stars,
we study the variation of the BP star counts with Galactic longitude and compare it
to those of a bona fide Milky Way and Monoceros population.
To this end, we have defined three regions in the CMD that should include predominantly 
disk, the old Monoceros TO and BP stars, respectively (see upper panel in \figref{comparative}). 
We then counted the number of stars in those three regions for 11 SDSS circular fields of 
$60'$ of radius between the coordinates $(l,b)=(180,25)^{\circ}$ and $(\ell,b)=(230,25)^{\circ}$, 
equidistantly spaced every $5^{\circ}$ in $\ell$. In the bottom panel of \figref{comparative} 
we show the gradients in the star counts.To facilitate the comparison between the different boxes the sequences have 
been normalized by their value at $\ell = 180^{\circ}$. The number of disk stars remains nearly constant for the range of $\ell$ considered while TO and BP counts show a similar behavior. The observed increase in the counts is consistent with the \cite{Penarrubia2005} model. This result supports the interpretation that the area in the CMD denoted as BP is not populated by stars associated with the Milky Way disk but by a population possibly associated with Monoceros.

In order to show the morphology of these populations in the CMD with higher contrast, we have compared the SDSS CMDs for those positions that, according to the bottom panel in \figref{comparative}, present a larger difference in density of stars belonging to the ring. In the left panel of \figref{hess}, we show the Hess diagram corresponding to the CMD obtained at $(\ell,b)=(210,25)^{\circ}$, when the Milky Way components are removed by subtraction of the CMD corresponding to $(\ell,b)=(180,25)^{\circ}$. The contour of minimal density (lighter grey) indicates the bins with a $4\%$ of the highest density observed. The stellar overdensity above the traditional Monoceros TO has a lower significance with respect to the main ring population but it clearly confirms the existence of an extra component brighter than the Monoceros TO. The right panel of \figref{hess} shows the cumulative luminosity function constructed from the Hess diagram for the color range $0.3<g-r<0.5$. The 
change of slope in the luminosity function at $g\sim19.5$ is a possible hint that we are detecting the TO of the older ($\sim9$\,Gyr) population, supporting the interpretation that there are multiple populations present in these fields, although we regard this result with caution.\\

\subsection{Comparison with synthetic Milky Way models}
\label{model_comparison}

  \begin{figure}[t]
     \begin{center}
      \includegraphics[scale=0.65]{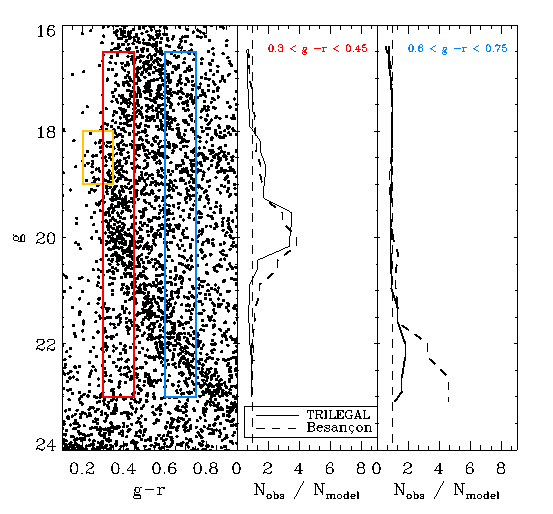}
      \caption[CMDs]{{\small \emph{Left:} CMD corresponding to the surrounding
       area of NGC\,2419. The red and blue vertical rectangles indicates the color ranges $0.3 < g-r <0.45$ and $0.6 < g-r <0.75$ respectively, used for the comparison with TRILEGAL and Besan\c con. The yellow rectangle shows the 
       position of the suggested BP. \emph{Right:} $N_{\rm obs} / N_{\rm model}$ as 
       a function of $g$, where $N_{\rm obs}$ and $N_{\rm model}$ are the counts 
       obtained for the observed and synthetic CMDs respectively for the color ranges considered. The solid and dashed lines indicate the results corresponding to TRILEGAL and Besan\c con.}}
\label{models}
     \end{center}
    \end{figure}

Since it is expected that the regions of the CMDs where Monoceros lies 
are also populated by Milky Way stars, we 
estimate the contribution of stars belonging to 
Galactic components by comparing the observed diagrams with synthetic CMDs 
generated with the Milky Way photometric models TRILEGAL and Besan\c con 
\citep{Robin2003,Girardi2005,Vanhollebeke2009} for the same line-of-sight to each cluster 
and for a similar solid angle. We have used the set of optimized parameters provided by \cite{Gao2013} as inputs for TRILEGAL while default values were used for Besan\c con.

We defined a narrow color range of $0.30< g-r <0.45$ in the CMD to be 
analyzed and compared with the TRILEGAL and Besan\c con counts (\figref{models}). We 
counted the number of stars in each of the 15 bins in which the $16.5< g <23$ 
range was divided and compared this number with that obtained for the same color/magnitude bin in the synthetic 
CMDs\footnote{When doing this comparison it was not necessary to correct 
the observed data for completeness since at the faint magnitude limit of the 
comparison our photometry is more than 90\% complete.}. 
In the middle panel of \figref{models} we show the fraction $N_{\rm obs} / N_{\rm model}$ 
as a function of $g$ magnitude, where $N_{\rm obs}$ and $N_{\rm model}$ are the counts 
obtained for the observed and synthetic CMDs respectively. With the exception 
of the range $18< g<21$, where the presence of Monoceros MS stars is 
visually dominant, the models considered here reproduce adequately the 
observed distribution for both the disk and halo stars. The area in the CMD 
where the contribution of Monoceros stars stands out from the expected 
stellar counts includes the area defined in this work as BP, however the significance is lower in the comparison with Besan\c con.  Similar results are obtained when we use the surroundings of Kop\,2 to carry out these tests.

\subsection{A younger population or metallicity spread?}
\label{young_population}

  \begin{figure}[h!]
     \begin{center}
      \includegraphics[scale=0.9]{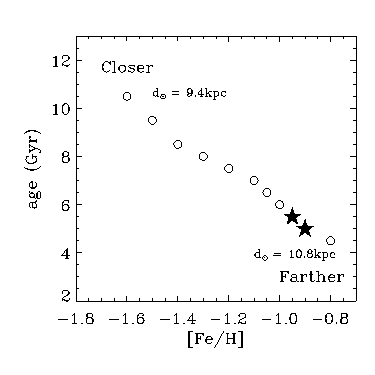}
      \caption[isochrones]{{\small Possible [Fe/H]--age values for the isochrones that are able to reproduce the BP, including the Monoceros MS up to its faintest end. The filled stars correspond to those combinations resulting in a heliocentric distance compatible with that derived for the Monoceros old population ($d_{\odot} \sim 10$\,kpc).}}
\label{possible_isochrones}
     \end{center}
    \end{figure}

 \begin{figure}[t]
     \begin{center}
      \includegraphics[scale=0.9]{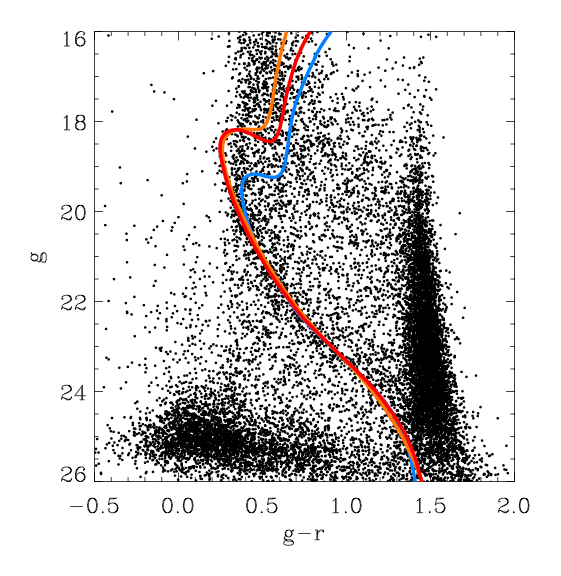}
      \caption[CMDs]{{\small CMD correspondings to the surroundings 
      of NGC\,2419. The blue isochrone represents the Monoceros 
      ring stellar population while the red isochrone indicates
      the $t \sim 5$\,Gyr and [Fe/H]$ \sim -0.95$ population at a similar heliocentric distance that of the halo substructure.The orange isochrone corresponds to a $t \sim 9$\,Gyr and [Fe/H]$ \sim -1.5$ population.}}
\label{cmds2}
     \end{center}
    \end{figure}

We have determined that the BP  corresponds to an actual overdensity and a differentiated population with respect to the Milky Way components. In order to characterize the BP, we used the same family of theoretical isochrones by 
\cite{Dotter2008} assuming different combinations of age and [Fe/H]. We considered a grid of possible isochrones with ages between $5-10$\,Gyr and $-1.6 < $[Fe/H]$ < -0.8$ to find the best matches to Monoceros (including the BP region). Reasonable matches are obtained within a range of ages and metallicities (see \figref{possible_isochrones}), resulting in distances in the range $9 < d_{\odot} < 12$\,kpc. However, the combination that yields a heliocentric distance similar to the one previously  derived for Monoceros is an isochrone corresponding to $t \sim 5$\,Gyr and [Fe/H]$ \sim -0.95$ (\figref{cmds2}). This shows that Monoceros might be composed of, at least, two stellar populations:  a dominant component of $\sim9$\,Gyr and a second contribution $\sim 4$\,Gyr younger. Note that our photometry, where the faintest part of the MS is clearly defined, allows us to discard other possible solutions that would only match the upper MS, a limitation when using shallower data as those from SDSS (see \figref{cmds}).

The presence of multiple stellar populations in streams is a natural consequence of the complex properties of their progenitor galaxies. In the case of the Sgr dwarf spheroidal, the presence of old, intermediate and young-aged star formation epochs has been clearly established \citep[e.g. ][]{Fahlman1996,Marconi1998,Bonifacio2004,Bellazzini2006a,Siegel2007}. In the case of Monoceros, the presence of M giant stars in that structure around the anticenter \citep{Rocha-Pinto2003} and the metallicity values reported along different lines-of-sight suggested that possibility. Our detection of a possible younger population in that same region of Monoceros might help establish whether the Monoceros ring is a complex halo substructure generated by the accretion of a minor satellite or by the distortion of the Galactic disk. 

From \figref{possible_isochrones}, we conclude that the BP might be also reproduced assuming the isochrone corresponding to a population with a similar age that of the main Monoceros population but with a lower metallicity of [Fe/H] $\sim -1.5$ (see \figref{cmds2}). This result agrees with the mean metallicity value obtained by \cite{Yanny2003} and the derived distance ($d_{\odot} \sim 9$\,kpc) is consistent within errors with the heliocentric distance for the main Monoceros MS population in that line-of-sight. Therefore, the presence of the BP component in the obtained CMDs might be produced by the manifestation of a metallicity spread suggested by the wide range of values reported for Monoceros in the literature. Alternatively, it is also possible that only a 5\,Gyr stellar population is present along this line-of-sight. However, the multiple detections of Monoceros in other areas of the sky have provided an age $t \sim 9$\,Gyr so this scenario might be difficult to reconcile with previous work focused on the ring. 

It would be interesting to confirm the presence in other sections of the Monoceros BP using deep wide-field photometry. Even the possible non-detection of hypothetical younger stars in other sections of Monoceros might help us better constrain the location of the progenitor system of this halo substructure, taking advantage of the fact that the different stellar populations are expected to leave the 
satellite main body at different times \citep{Penarrubia2008,Walker2011}.

\subsection{Blue stragglers}

Since the BP is located blueward and above the TO position of Monoceros, it populates the area of the CMD where blue straggler stars are expected. Recently, \cite{Santana2013} have shown that blue stragglers are ubiquitous among Galactic GCs, classical dwarf spheroidal and ultra-faint dwarf galaxies. Their analysis show that to reproduce the observed number of blue stragglers  in dwarf galaxies, this population should be composed of stars with age $t\sim2.5$\,Gyr and account for up to the unlikely fined-tuned fraction of $7\%$ of the total number of stars in the satellite. Despite the fact that the distance uncertainties and depth along the line-of-sight of Monoceros do not allow us to  estimate a precise age for the BP, we do not see a significant population of stars extending as far as the MSTO of a $2.5$\,Gyr old population given by the \cite{Dotter2008} theoretical isochrones. That level is estimated assuming that the brightest members of a blue straggler population should reach a magnitude of $g_{\rm TO} \sim 16.3$, corresponding to a stellar mass of twice the mass of a Monoceros MSTO star, at least $\sim 2$ magnitudes brighter than the BP MSTO.

In addition, from our data we can estimate the specific fraction of blue straggler stars if we assume that the BP is populated entirely by them. Following the procedure described in previous work \citep[e.g.][]{Sollima2008a,Santana2013}, we selected all those stars in the BP and along the Monoceros MS in the range $20.5 < g < 21.5$. After the decontamination of the observed stellar counts using as reference the synthetic CMD generated with the Besan\c con model for \secref{model_comparison}, we conclude that a BP composed of blue stragglers would imply a specific fraction of $F = 20\%$, an order of magnitude larger than the observed fraction
of blue stragglers in globular clusters and dwarf galaxies. Therefore, we consider the possibility of the BP being made up entirely of blue
straggler stars as unlikely.

\subsection{Foreground population}
\label{second_wrap}

So far, we have shown that the BP is likely not associated with a Milky Way disk population and that it is consistent with being part of Monoceros. However, it is still possible that the BP corresponds to a population different from both the Milky Way and Monoceros or to a different wrap of that halo substructure. Given the appearance of the BP in the CMDs, a possibility is that it corresponds to a feature along the line-of-sight between us and Monoceros. If this was the case, we should be able to detect its presence at fainter magnitudes and redder colors, slightly above the well-defined Monoceros MS. To check this scenario, the same comparison with the Milky Way models as in \secref{model_comparison} is performed, but this time in the color range $0.6< g-r <0.75$. The right panel in \figref{models} shows the results. 

In this case both models reproduce the observed counts with the exception of the $g > 21$ range, where we find around four times more stars presumably associated with Monoceros than in the synthetic CMD generated with Besan\c con. The presence of more than one subjacent MS in the redder region of the CMD is not obvious, so there are not significant evidence in our data of a foreground stream possibly associated with the BP feature.

According to the \cite{Penarrubia2005} model (\figref{monoceros_model}), two different wraps of Monoceros might be present at different distances along the direction to NGC\,2419 and Kop\,2 and with indiscernible velocities (see discussion below). Therefore, our results are compatible with the detection of a different wrap of the ring in that area of the sky although the metallicity of such component would differ from that of the main population of Monoceros observed in these fields (\figref{possible_isochrones}). This might support the scenario in which the generation of the BP is related to the hypothetical Monoceros metallicity spread.

\subsection{Spectroscopic confirmation of second population in the line-of-sight}

 \begin{figure}[t]
     \begin{center}
      \includegraphics[scale=0.9]{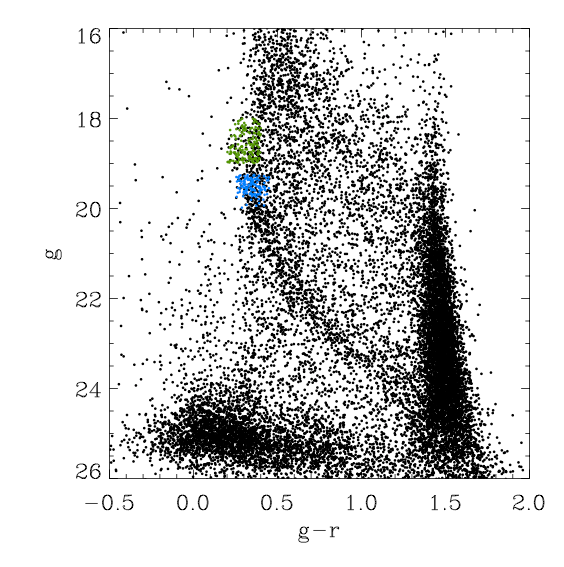}
      \caption[CMDs]{{\small CMD correspondings to the surroundings 
      of NGC\,2419. The green and blue points correspond to the position of the BP and Monoceros TO stars used for the analysis of the SDSS spectroscopy data, respectively. }}
\label{cmds_espec}
     \end{center}
    \end{figure}

\begin{figure*}[!t]
     \begin{center}
      \includegraphics[scale=0.35]{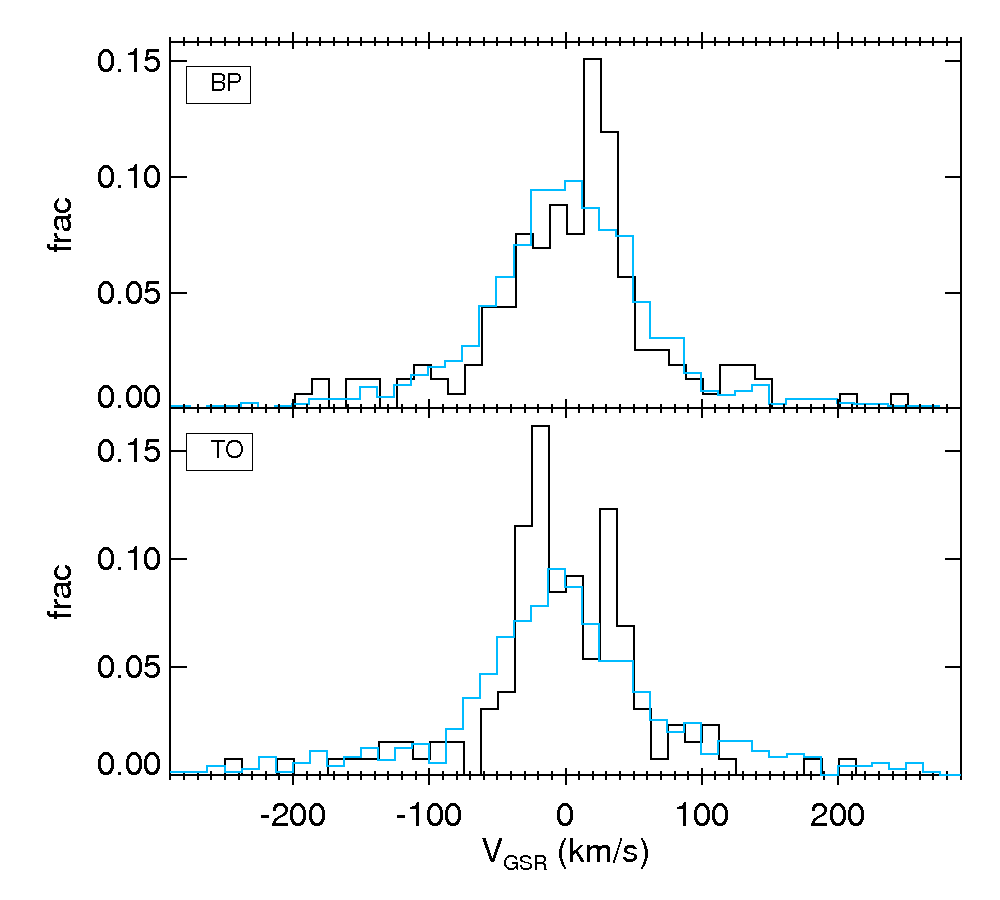}
      \includegraphics[scale=0.35]{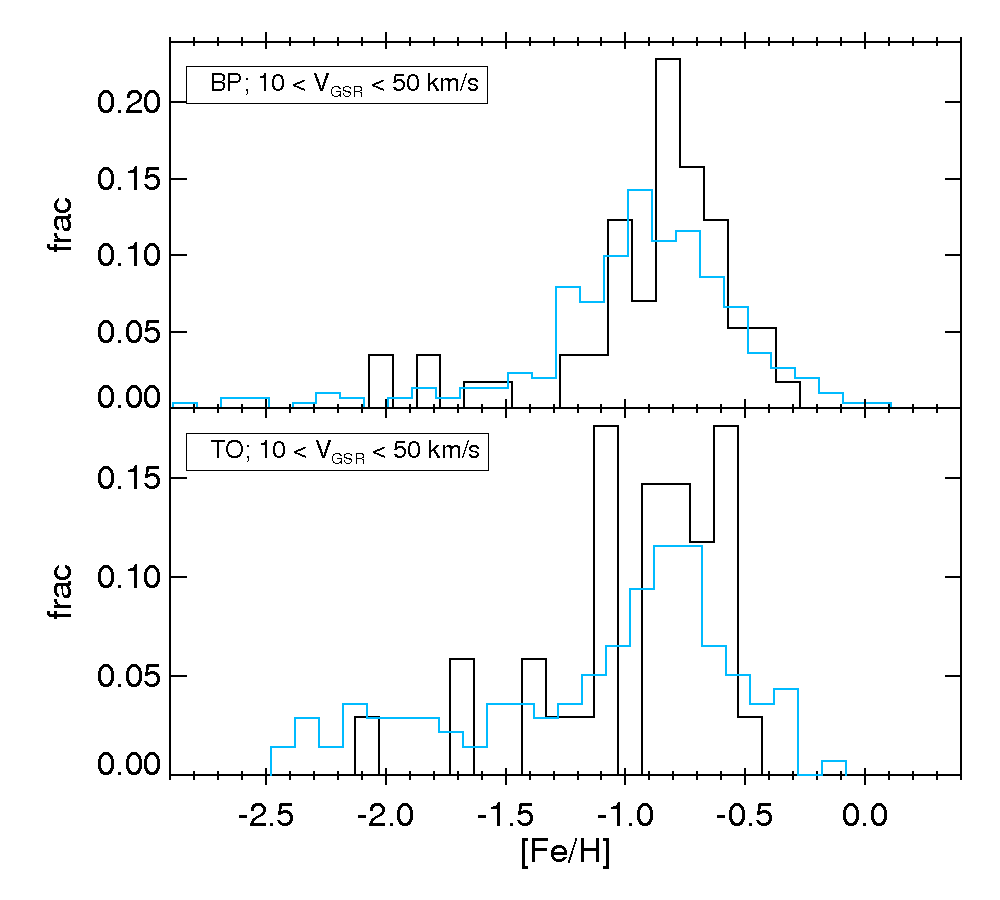}
      \caption[CMDs]{{\small Spectroscopic data from SDSS DR10 for the region near NGC 2419 (specifically, $175\arcdeg < l < 184\arcdeg$, $22.7\arcdeg < b < 27.7\arcdeg$). The upper and lower panels depict velocities (left panels) and metallicities (right panels) of stars with colors $0.2 < g-r < 0.4$ in the magnitude ranges $18.0 < g < 19.0$ for the BP population and $0.25 < g-r < 0.45$ and $19.25 < g < 20.0$ for the TO population. A kinematically cold peak is clearly visible at $V_{\rm gsr} \sim 20$~km~s$^{-1}$ in both the BP and TO samples. The right panels show metallicities for only stars between $0 < V_{\rm gsr} < 50$~km~s$^{-1}$. The TO and BP stars present a similar metallicity.
 }}
\label{n2419_spect}
     \end{center}
    \end{figure*}

To verify our interpretation that the BP represents a younger sub-population at the same distance as the Monoceros structure, we turn to spectroscopy from SDSS DR10 \citep{ahn2014}. We select 160 stars around the position of NGC\,2419 with colors $0.2 < g-r < 0.4$ in the magnitude ranges $18.0 < g < 19.0$ for the BP population and 131 stars with $0.25 < g-r < 0.45$ and $19.25 < g < 20.0$ for the TO population (see \figref{cmds_espec}). For comparison, we also generate Besan\c con models in the same field of view, and select stars with identical color-magnitude criteria. Figure~\ref{n2419_spect} shows the resulting velocities (left panels) and metallicities (right) for the BP (upper panels) and TO (lower) in this region. Model predictions are given as blue histograms. A clear excess population is seen in both the BP and TO populations at $V_{\rm gsr} \sim 20$~km~s$^{-1}$. The velocity dispersion of this structure is clearly much less than the thick disk (which constitutes most of the Besancon stars in the BP selection box), confirming the presence of a kinematically cold structure in both the BP and TO samples, with a velocity compatible with that of the Monoceros ring in that line-of-sight as predicted by the \cite{Penarrubia2005} model (see bottom panel in \figref{monoceros_model}). These data are selected in the same sky area as fields denoted ``M3'' and ``M4'' by \citet{Li2012}, who found a narrow excess at $V_{\rm gsr} \sim 30$~km~s$^{-1}$ in field M3 which is likely the same population we see in Figure~\ref{n2419_spect}.

In the right panels of \figref{n2419_spect}, we show the metallicities of stars from each population with $0 < V_{\rm gsr} < 50$ km s$^{-1}$ to emphasize the velocity substructure. The mean metallicities of the BP and TO populations in this velocity range are similar, confirming our finding based on isochrone fitting in \secref{young_population}.
\newpage

\section{Conclusions}

We have used the deep wide-field photometry obtained for the GCs NGC\,2419 and Kop\,2 to investigate the presence of a second and younger stellar population in the Monoceros ring. Using SDSS photometry and spectroscopy and the Milky Way synthetic model TRILEGAL, we conclude that there exists a differentiated stellar population which is not associated with any of the other Galactic components and that we have denoted as BP. That population also stands out when we compare with Besan\c con, a synthetic model which includes a disk warp and flare.

Isochrone fitting shows that one of the possible explanations for the presence of BP in the CMDs is that Monoceros is composed of an old MS bulk of stars and an additional second population $\sim 4$\,Gyr younger, with similar metallicity and lying at a heliocentric distance of $d_{\odot} \sim10$\,kpc. We have confirmed, using the radial velocities provided by SDSS spectroscopy, that the proposed younger population presents a similar kinematics to that of the stellar ring. Alternatively, a metallicity spread might generate a feature as the one observed in the CMD. These results suggests that the hypothetical progenitor galaxy that generated the Monoceros ring after its accretion might present a more complex composition.

On the other hand, our analysis suggests that a second foreground stellar system along the same line-of-sight might also reproduce the observed CMD morphology. According to the predicted distribution of Monoceros ring tidal debris, it is possible that this detection corresponds to a second wrap of that substructure and with a different metallicity. 

Further deep wide-field photometry of other areas of the sky with high density of Monoceros ring stars and at intermediate Galactic latitudes is needed to establish the true nature of the population unveiled in the direction of these two GCs. All these evidence might shed light on the origin of this controversial halo stellar overdensity.

\acknowledgments
We warmly thank the anonymous referees for their helpful comments and suggestions. J.~A.~C-B and R.~R.~M. received support from Centre of Excellence in Astrophysics and Associated Technologies (PFB-06). R.~R.~M.~acknowledges partial support from CONICYT Anillo project ACT-1122 as well as FONDECYT project N$^{\circ}1120013$. JLC acknowledges support by NSF grant AST 09-37523. SGD ackowledges a partial support from the NSF grants AST-0909182, AST-1313422 and AST-1413600. 

\bibliographystyle{apj}
\bibliography{bibliomon}{}

\begin{thebibliography}{62}
\expandafter\ifx\csname natexlab\endcsname\relax\def\natexlab#1{#1}\fi

\bibitem[{{Ahn} {et~al.}(2014){Ahn}, {Alexandroff}, {Allende Prieto}, {Anders},
  {Anderson}, {Anderton}, {Andrews}, {Aubourg}, {Bailey}, {Bastien}, \&
  et~al.}]{ahn2014}
{Ahn}, C.~P., {Alexandroff}, R., {Allende Prieto}, C., {et~al.} 2014, \apjs,
  211, 17

\bibitem[{{Bellazzini} {et~al.}(2006){Bellazzini}, {Ibata}, {Martin}, {Lewis},
  {Conn}, \& {Irwin}}]{Bellazzini2006a}
{Bellazzini}, M., {Ibata}, R., {Martin}, N., {et~al.} 2006, \mnras, 366, 865

\bibitem[{{Belokurov} {et~al.}(2006){Belokurov}, {Zucker}, {Evans}, {Gilmore},
  {Vidrih}, {Bramich}, {Newberg}, {Wyse}, {Irwin}, {Fellhauer}, {Hewett},
  {Walton}, {Wilkinson}, {Cole}, {Yanny}, {Rockosi}, {Beers}, {Bell},
  {Brinkmann}, {Ivezi{\'c}}, \& {Lupton}}]{Belokurov2006}
{Belokurov}, V., {Zucker}, D.~B., {Evans}, N.~W., {et~al.} 2006, \apjl, 642,
  L137

\bibitem[{{Belokurov} {et~al.}(2007){Belokurov}, {Evans}, {Bell}, {Irwin},
  {Hewett}, {Koposov}, {Rockosi}, {Gilmore}, {Zucker}, {Fellhauer},
  {Wilkinson}, {Bramich}, {Vidrih}, {Rix}, {Beers}, {Schneider}, {Barentine},
  {Brewington}, {Brinkmann}, {Harvanek}, {Krzesinski}, {Long}, {Pan},
  {Snedden}, {Malanushenko}, \& {Malanushenko}}]{Belokurov2007a}
{Belokurov}, V., {Evans}, N.~W., {Bell}, E.~F., {et~al.} 2007, \apjl, 657, L89

\bibitem[{{Bonifacio} {et~al.}(2004){Bonifacio}, {Sbordone}, {Marconi},
  {Pasquini}, \& {Hill}}]{Bonifacio2004}
{Bonifacio}, P., {Sbordone}, L., {Marconi}, G., {Pasquini}, L., \& {Hill}, V.
  2004, \aap, 414, 503

\bibitem[{{Butler} {et~al.}(2007){Butler}, {Mart{\'{\i}}nez-Delgado}, {Rix},
  {Pe{\~n}arrubia}, \& {de Jong}}]{Butler2007}
{Butler}, D.~J., {Mart{\'{\i}}nez-Delgado}, D., {Rix}, H.-W., {Pe{\~n}arrubia},
  J., \& {de Jong}, J.~T.~A. 2007, \aj, 133, 2274

\bibitem[{{Carballo-Bello} {et~al.}(2012){Carballo-Bello}, {Gieles}, {Sollima},
  {Koposov}, {Mart{\'{\i}}nez-Delgado}, \&
  {Pe{\~n}arrubia}}]{Carballo-Bello2012}
{Carballo-Bello}, J.~A., {Gieles}, M., {Sollima}, A., {et~al.} 2012, \mnras,
  419, 14

\bibitem[{{Carballo-Bello} {et~al.}(2014){Carballo-Bello}, {Sollima},
  {Martinez-Delgado}, {Pila-Diez}, {Leaman}, {Fliri}, {Munoz}, \&
  {Corral-Santana}}]{Carballo-Bello2014}
{Carballo-Bello}, J.~A., {Sollima}, A., {Martinez-Delgado}, D., {et~al.} 2014,
  ArXiv:1409.7390

\bibitem[{{Chou} {et~al.}(2010){Chou}, {Cunha}, {Majewski}, {Smith},
  {Patterson}, {Mart{\'{\i}}nez-Delgado}, \& {Geisler}}]{Chou2010}
{Chou}, M.-Y., {Cunha}, K., {Majewski}, S.~R., {et~al.} 2010, \apj, 708, 1290

\bibitem[{{Conn} {et~al.}(2008){Conn}, {Lane}, {Lewis}, {Irwin}, {Ibata},
  {Martin}, {Bellazzini}, \& {Tuntsov}}]{Conn2008}
{Conn}, B.~C., {Lane}, R.~R., {Lewis}, G.~F., {et~al.} 2008, \mnras, 390, 1388

\bibitem[{{Conn} {et~al.}(2005){Conn}, {Lewis}, {Irwin}, {Ibata}, {Ferguson},
  {Tanvir}, \& {Irwin}}]{Conn2005}
{Conn}, B.~C., {Lewis}, G.~F., {Irwin}, M.~J., {et~al.} 2005, \mnras, 362, 475

\bibitem[{{Conn} {et~al.}(2007){Conn}, {Lane}, {Lewis}, {Gil-Merino}, {Irwin},
  {Ibata}, {Martin}, {Bellazzini}, {Sharp}, {Tuntsov}, \&
  {Ferguson}}]{Conn2007}
{Conn}, B.~C., {Lane}, R.~R., {Lewis}, G.~F., {et~al.} 2007, \mnras, 376, 939

\bibitem[{{Conn} {et~al.}(2012){Conn}, {No{\"e}l}, {Rix}, {Lane}, {Lewis},
  {Irwin}, {Martin}, {Ibata}, {Dolphin}, \& {Chapman}}]{Conn2012}
{Conn}, B.~C., {No{\"e}l}, N.~E.~D., {Rix}, H.-W., {et~al.} 2012, \apj, 754,
  101

\bibitem[{{Crane} {et~al.}(2003){Crane}, {Majewski}, {Rocha-Pinto},
  {Frinchaboy}, {Skrutskie}, \& {Law}}]{Crane2003}
{Crane}, J.~D., {Majewski}, S.~R., {Rocha-Pinto}, H.~J., {et~al.} 2003, \apjl,
  594, L119

\bibitem[{{Deason} {et~al.}(2014){Deason}, {Belokurov}, {Koposov}, \&
  {Rockosi}}]{Deason2014}
{Deason}, A.~J., {Belokurov}, V., {Koposov}, S.~E., \& {Rockosi}, C.~M. 2014,
  \apj, 787, 30

\bibitem[{{Dotter} {et~al.}(2008){Dotter}, {Chaboyer}, {Jevremovi{\'c}},
  {Kostov}, {Baron}, \& {Ferguson}}]{Dotter2008}
{Dotter}, A., {Chaboyer}, B., {Jevremovi{\'c}}, D., {et~al.} 2008, \apjs, 178,
  89

\bibitem[{{Fahlman} {et~al.}(1996){Fahlman}, {Mandushev}, {Richer}, {Thompson},
  \& {Sivaramakrishnan}}]{Fahlman1996}
{Fahlman}, G.~G., {Mandushev}, G., {Richer}, H.~B., {Thompson}, I.~B., \&
  {Sivaramakrishnan}, A. 1996, \apjl, 459, L65+

\bibitem[{{Gao} {et~al.}(2013){Gao}, {Just}, \& {Grebel}}]{Gao2013}
{Gao}, S., {Just}, A., \& {Grebel}, E.~K. 2013, \aap, 549, A20

\bibitem[{{Girardi} {et~al.}(2005){Girardi}, {Groenewegen}, {Hatziminaoglou},
  \& {da Costa}}]{Girardi2005}
{Girardi}, L., {Groenewegen}, M.~A.~T., {Hatziminaoglou}, E., \& {da Costa}, L.
  2005, \aap, 436, 895

\bibitem[{{Grillmair}(2006)}]{Grillmair2006d}
{Grillmair}, C.~J. 2006, \apjl, 651, L29

\bibitem[{{Grillmair} {et~al.}(2008){Grillmair}, {Carlin}, \&
  {Majewski}}]{Grillmair2008}
{Grillmair}, C.~J., {Carlin}, J.~L., \& {Majewski}, S.~R. 2008, \apjl, 689,
  L117

\bibitem[{{Hammersley} \& {L{\'o}pez-Corredoira}(2011)}]{Hammersley2011}
{Hammersley}, P.~L., \& {L{\'o}pez-Corredoira}, M. 2011, \aap, 527, A6+

\bibitem[{{Harris}(2010)}]{Harris2010}
{Harris}, W.~E. 2010, ArXiv:1012.3224

\bibitem[{{Ibata} {et~al.}(1994){Ibata}, {Gilmore}, \& {Irwin}}]{Ibata1994}
{Ibata}, R.~A., {Gilmore}, G., \& {Irwin}, M.~J. 1994, \nat, 370, 194

\bibitem[{{Ivezi{\'c}} {et~al.}(2008){Ivezi{\'c}}, {Sesar}, {Juri{\'c}},
  {Bond}, {Dalcanton}, {Rockosi}, {Yanny}, {Newberg}, {Beers}, {Allende
  Prieto}, {Wilhelm}, {Lee}, {Sivarani}, {Norris}, {Bailer-Jones}, {Re
  Fiorentin}, {Schlegel}, {Uomoto}, {Lupton}, {Knapp}, {Gunn}, {Covey},
  {Smith}, {Miknaitis}, {Doi}, {Tanaka}, {Fukugita}, {Kent}, {Finkbeiner},
  {Munn}, {Pier}, {Quinn}, {Hawley}, {Anderson}, {Kiuchi}, {Chen}, {Bushong},
  {Sohi}, {Haggard}, {Kimball}, {Barentine}, {Brewington}, {Harvanek},
  {Kleinman}, {Krzesinski}, {Long}, {Nitta}, {Snedden}, {Lee}, {Harris},
  {Brinkmann}, {Schneider}, \& {York}}]{Ivezic2008}
{Ivezi{\'c}}, {\v Z}., {Sesar}, B., {Juri{\'c}}, M., {et~al.} 2008, \apj, 684,
  287

\bibitem[{{Juri{\'c}} {et~al.}(2008){Juri{\'c}}, {Ivezi{\'c}}, {Brooks},
  {Lupton}, {Schlegel}, {Finkbeiner}, {Padmanabhan}, {Bond}, {Sesar},
  {Rockosi}, {Knapp}, {Gunn}, {Sumi}, {Schneider}, {Barentine}, {Brewington},
  {Brinkmann}, {Fukugita}, {Harvanek}, {Kleinman}, {Krzesinski}, {Long},
  {Neilsen}, {Nitta}, {Snedden}, \& {York}}]{Juric2008}
{Juri{\'c}}, M., {Ivezi{\'c}}, {\v Z}., {Brooks}, A., {et~al.} 2008, \apj, 673,
  864

\bibitem[{{Koposov} {et~al.}(2007){Koposov}, {de Jong}, {Belokurov}, {Rix},
  {Zucker}, {Evans}, {Gilmore}, {Irwin}, \& {Bell}}]{Koposov2007}
{Koposov}, S., {de Jong}, J.~T.~A., {Belokurov}, V., {et~al.} 2007, \apj, 669,
  337

\bibitem[{{Koposov} {et~al.}(2012){Koposov}, {Belokurov}, {Evans}, {Gilmore},
  {Gieles}, {Irwin}, {Lewis}, {Niederste-Ostholt}, {Pe{\~n}arrubia}, {Smith},
  {Bizyaev}, {Malanushenko}, {Malanushenko}, {Schneider}, \&
  {Wyse}}]{Koposov2012}
{Koposov}, S.~E., {Belokurov}, V., {Evans}, N.~W., {et~al.} 2012, \apj, 750, 80

\bibitem[{{Li} {et~al.}(2012){Li}, {Newberg}, {Carlin}, {Deng}, {Newby},
  {Willett}, {Xu}, \& {Luo}}]{Li2012}
{Li}, J., {Newberg}, H.~J., {Carlin}, J.~L., {et~al.} 2012, \apj, 757, 151

\bibitem[{{Majewski} {et~al.}(2003){Majewski}, {Skrutskie}, {Weinberg}, \&
  {Ostheimer}}]{Majewski2003}
{Majewski}, S.~R., {Skrutskie}, M.~F., {Weinberg}, M.~D., \& {Ostheimer}, J.~C.
  2003, \apj, 599, 1082

\bibitem[{{Marconi} {et~al.}(1998){Marconi}, {Buonanno}, {Castellani},
  {Iannicola}, {Molaro}, {Pasquini}, \& {Pulone}}]{Marconi1998}
{Marconi}, G., {Buonanno}, R., {Castellani}, M., {et~al.} 1998, \aap, 330, 453

\bibitem[{{Martin} {et~al.}(2004){Martin}, {Ibata}, {Bellazzini}, {Irwin},
  {Lewis}, \& {Dehnen}}]{Martin2004}
{Martin}, N.~F., {Ibata}, R.~A., {Bellazzini}, M., {et~al.} 2004, \mnras, 348,
  12

\bibitem[{{Mart{\'{\i}}nez-Delgado} {et~al.}(2005){Mart{\'{\i}}nez-Delgado},
  {Butler}, {Rix}, {Franco}, {Pe{\~n}arrubia}, {Alfaro}, \&
  {Dinescu}}]{Martinez-Delgado2005}
{Mart{\'{\i}}nez-Delgado}, D., {Butler}, D.~J., {Rix}, H.-W., {et~al.} 2005,
  \apj, 633, 205

\bibitem[{{Mart{\'{\i}}nez-Delgado} {et~al.}(2004){Mart{\'{\i}}nez-Delgado},
  {Dinescu}, {Zinn}, {Tutsoff}, {C{\^o}t{\'e}}, \&
  {Boyarchuck}}]{MartinezDelgado2004}
{Mart{\'{\i}}nez-Delgado}, D., {Dinescu}, D.~I., {Zinn}, R., {et~al.} 2004, in
  Astronomical Society of the Pacific Conference Series, Vol. 327, Satellites
  and Tidal Streams, ed. {F.~Prada, D.~Martinez Delgado, \& T.~J.~Mahoney},
  255--+

\bibitem[{{Mateu} {et~al.}(2009){Mateu}, {Vivas}, {Zinn}, {Miller}, \&
  {Abad}}]{Mateu2009}
{Mateu}, C., {Vivas}, A.~K., {Zinn}, R., {Miller}, L.~R., \& {Abad}, C. 2009,
  \aj, 137, 4412

\bibitem[{{Meisner} {et~al.}(2012){Meisner}, {Frebel}, {Juri{\'c}}, \&
  {Finkbeiner}}]{Meisner2012}
{Meisner}, A.~M., {Frebel}, A., {Juri{\'c}}, M., \& {Finkbeiner}, D.~P. 2012,
  \apj, 753, 116

\bibitem[{{Michel-Dansac} {et~al.}(2011){Michel-Dansac}, {Abadi}, {Navarro}, \&
  {Steinmetz}}]{Michel-Dansac2011}
{Michel-Dansac}, L., {Abadi}, M.~G., {Navarro}, J.~F., \& {Steinmetz}, M. 2011,
  \mnras, 414, L1

\bibitem[{{Moitinho} {et~al.}(2006){Moitinho}, {V{\'a}zquez}, {Carraro},
  {Baume}, {Giorgi}, \& {Lyra}}]{Moitinho2006}
{Moitinho}, A., {V{\'a}zquez}, R.~A., {Carraro}, G., {et~al.} 2006, \mnras,
  368, L77

\bibitem[{{Momany} {et~al.}(2006){Momany}, {Zaggia}, {Gilmore}, {Piotto},
  {Carraro}, {Bedin}, \& {de Angeli}}]{Momany2006}
{Momany}, Y., {Zaggia}, S., {Gilmore}, G., {et~al.} 2006, \aap, 451, 515

\bibitem[{{Momany} {et~al.}(2004){Momany}, {Zaggia}, {Bonifacio}, {Piotto}, {De
  Angeli}, {Bedin}, \& {Carraro}}]{Momany2004}
{Momany}, Y., {Zaggia}, S.~R., {Bonifacio}, P., {et~al.} 2004, \aap, 421, L29

\bibitem[{{Mu{\~n}oz} {et~al.}(2010){Mu{\~n}oz}, {Geha}, \&
  {Willman}}]{Munoz2010}
{Mu{\~n}oz}, R.~R., {Geha}, M., \& {Willman}, B. 2010, \aj, 140, 138

\bibitem[{{Natarajan} \& {Sikivie}(2007)}]{Natarajan2007}
{Natarajan}, A., \& {Sikivie}, P. 2007, \prd, 76, 023505

\bibitem[{{Newberg} {et~al.}(2002){Newberg}, {Yanny}, {Rockosi}, {Grebel},
  {Rix}, {Brinkmann}, {Csabai}, {Hennessy}, {Hindsley}, {Ibata}, {Ivezi{\'c}},
  {Lamb}, {Nash}, {Odenkirchen}, {Rave}, {Schneider}, {Smith}, {Stolte}, \&
  {York}}]{Newberg2002}
{Newberg}, H.~J., {Yanny}, B., {Rockosi}, C., {et~al.} 2002, \apj, 569, 245

\bibitem[{{Pe{\~n}arrubia} {et~al.}(2008){Pe{\~n}arrubia}, {Navarro}, \&
  {McConnachie}}]{Penarrubia2008}
{Pe{\~n}arrubia}, J., {Navarro}, J.~F., \& {McConnachie}, A.~W. 2008, \apj,
  673, 226

\bibitem[{{Pe{\~n}arrubia} {et~al.}(2005){Pe{\~n}arrubia},
  {Mart{\'{\i}}nez-Delgado}, {Rix}, {G{\'o}mez-Flechoso}, {Munn}, {Newberg},
  {Bell}, {Yanny}, {Zucker}, \& {Grebel}}]{Penarrubia2005}
{Pe{\~n}arrubia}, J., {Mart{\'{\i}}nez-Delgado}, D., {Rix}, H.~W., {et~al.}
  2005, \apj, 626, 128

\bibitem[{{Purcell} {et~al.}(2011){Purcell}, {Bullock}, {Tollerud}, {Rocha}, \&
  {Chakrabarti}}]{Purcell2011}
{Purcell}, C.~W., {Bullock}, J.~S., {Tollerud}, E.~J., {Rocha}, M., \&
  {Chakrabarti}, S. 2011, \nat, 477, 301

\bibitem[{{Ripepi} {et~al.}(2007){Ripepi}, {Clementini}, {Di Criscienzo},
  {Greco}, {Dall'Ora}, {Federici}, {Di Fabrizio}, {Musella}, {Marconi},
  {Baldacci}, \& {Maio}}]{Ripepi2007}
{Ripepi}, V., {Clementini}, G., {Di Criscienzo}, M., {et~al.} 2007, \apjl, 667,
  L61

\bibitem[{{Robin} {et~al.}(2003){Robin}, {Reyl{\'e}}, {Derri{\`e}re}, \&
  {Picaud}}]{Robin2003}
{Robin}, A.~C., {Reyl{\'e}}, C., {Derri{\`e}re}, S., \& {Picaud}, S. 2003,
  \aap, 409, 523

\bibitem[{{Rocha-Pinto} {et~al.}(2003){Rocha-Pinto}, {Majewski}, {Skrutskie},
  \& {Crane}}]{Rocha-Pinto2003}
{Rocha-Pinto}, H.~J., {Majewski}, S.~R., {Skrutskie}, M.~F., \& {Crane}, J.~D.
  2003, \apjl, 594, L115

\bibitem[{{Santana} {et~al.}(2013){Santana}, {Mu{\~n}oz}, {Geha},
  {C{\^o}t{\'e}}, {Stetson}, {Simon}, \& {Djorgovski}}]{Santana2013}
{Santana}, F.~A., {Mu{\~n}oz}, R.~R., {Geha}, M., {et~al.} 2013, \apj, 774, 106

\bibitem[{{Schlafly} \& {Finkbeiner}(2011)}]{Schlafly2011}
{Schlafly}, E.~F., \& {Finkbeiner}, D.~P. 2011, \apj, 737, 103

\bibitem[{{Siegel} {et~al.}(2007){Siegel}, {Dotter}, {Majewski}, {Sarajedini},
  {Chaboyer}, {Nidever}, {Anderson}, {Mar{\'{\i}}n-Franch}, {Rosenberg},
  {Bedin}, {Aparicio}, {King}, {Piotto}, \& {Reid}}]{Siegel2007}
{Siegel}, M.~H., {Dotter}, A., {Majewski}, S.~R., {et~al.} 2007, \apjl, 667,
  L57

\bibitem[{{Simion} {et~al.}(2014){Simion}, {Belokurov}, {Irwin}, \&
  {Koposov}}]{Simion2014}
{Simion}, I.~T., {Belokurov}, V., {Irwin}, M., \& {Koposov}, S.~E. 2014,
  \mnras, 440, 161

\bibitem[{{Slater} {et~al.}(2014){Slater}, {Bell}, {Schlafly}, {Morganson},
  {Martin}, {Rix}, {Pe{\~n}arrubia}, {Bernard}, {Ferguson}, {Martinez-Delgado},
  {Wyse}, {Burgett}, {Chambers}, {Draper}, {Hodapp}, {Kaiser}, {Magnier},
  {Metcalfe}, {Price}, {Tonry}, {Wainscoat}, \& {Waters}}]{Slater2014}
{Slater}, C.~T., {Bell}, E.~F., {Schlafly}, E.~F., {et~al.} 2014, \apj, 791, 9

\bibitem[{{Sollima} {et~al.}(2008){Sollima}, {Lanzoni}, {Beccari}, {Ferraro},
  \& {Fusi Pecci}}]{Sollima2008a}
{Sollima}, A., {Lanzoni}, B., {Beccari}, G., {Ferraro}, F.~R., \& {Fusi Pecci},
  F. 2008, \aap, 481, 701

\bibitem[{{Sollima} {et~al.}(2011){Sollima}, {Valls-Gabaud},
  {Martinez-Delgado}, {Fliri}, {Pe{\~n}arrubia}, \& {Hoekstra}}]{Sollima2011}
{Sollima}, A., {Valls-Gabaud}, D., {Martinez-Delgado}, D., {et~al.} 2011,
  \apjl, 730, L6+

\bibitem[{{Stetson}(1994)}]{Stetson1994}
{Stetson}, P.~B. 1994, \pasp, 106, 250

\bibitem[{{Vanhollebeke} {et~al.}(2009){Vanhollebeke}, {Groenewegen}, \&
  {Girardi}}]{Vanhollebeke2009}
{Vanhollebeke}, E., {Groenewegen}, M.~A.~T., \& {Girardi}, L. 2009, \aap, 498,
  95

\bibitem[{{Walker et al.}(2011)}]{Walker2011}
{Walker et al.} 2011, \mnras, 415, 643

\bibitem[{{Yanny} {et~al.}(2003){Yanny}, {Newberg}, {Grebel}, {Kent},
  {Odenkirchen}, {Rockosi}, {Schlegel}, {Subbarao}, {Brinkmann}, {Fukugita},
  {Ivezic}, {Lamb}, {Schneider}, \& {York}}]{Yanny2003}
{Yanny}, B., {Newberg}, H.~J., {Grebel}, E.~K., {et~al.} 2003, \apj, 588, 824

\bibitem[{{York} {et~al.}(2000){York}, {Adelman}, {Anderson}, {Anderson},
  {Annis}, {Bahcall}, {Bakken}, {Barkhouser}, {Bastian}, {Berman}, {Boroski},
  {Bracker}, {Briegel}, {Briggs}, {Brinkmann}, {Brunner}, {Burles}, {Carey},
  {Carr}, {Castander}, {Chen}, {Colestock}, {Connolly}, {Crocker}, {Csabai},
  {Czarapata}, {Davis}, {Doi}, {Dombeck}, {Eisenstein}, {Ellman}, {Elms},
  {Evans}, {Fan}, {Federwitz}, {Fiscelli}, {Friedman}, {Frieman}, {Fukugita},
  {Gillespie}, {Gunn}, {Gurbani}, {de Haas}, {Haldeman}, {Harris}, {Hayes},
  {Heckman}, {Hennessy}, {Hindsley}, {Holm}, {Holmgren}, {Huang}, {Hull},
  {Husby}, {Ichikawa}, {Ichikawa}, {Ivezi{\'c}}, {Kent}, {Kim}, {Kinney},
  {Klaene}, {Kleinman}, {Kleinman}, {Knapp}, {Korienek}, {Kron}, {Kunszt},
  {Lamb}, {Lee}, {Leger}, {Limmongkol}, {Lindenmeyer}, {Long}, {Loomis},
  {Loveday}, {Lucinio}, {Lupton}, {MacKinnon}, {Mannery}, {Mantsch}, {Margon},
  {McGehee}, {McKay}, {Meiksin}, {Merelli}, {Monet}, {Munn}, {Narayanan},
  {Nash}, {Neilsen}, {Neswold}, {Newberg}, {Nichol}, {Nicinski}, {Nonino},
  {Okada}, {Okamura}, {Ostriker}, {Owen}, {Pauls}, {Peoples}, {Peterson},
  {Petravick}, {Pier}, {Pope}, {Pordes}, {Prosapio}, {Rechenmacher}, {Quinn},
  {Richards}, {Richmond}, {Rivetta}, {Rockosi}, {Ruthmansdorfer}, {Sandford},
  {Schlegel}, {Schneider}, {Sekiguchi}, {Sergey}, {Shimasaku}, {Siegmund},
  {Smee}, {Smith}, {Snedden}, {Stone}, {Stoughton}, {Strauss}, {Stubbs},
  {SubbaRao}, {Szalay}, {Szapudi}, {Szokoly}, {Thakar}, {Tremonti}, {Tucker},
  {Uomoto}, {Vanden Berk}, {Vogeley}, {Waddell}, {Wang}, {Watanabe},
  {Weinberg}, {Yanny}, \& {Yasuda}}]{York2000}
{York}, D.~G., {Adelman}, J., {Anderson}, Jr., J.~E., {et~al.} 2000, \aj, 120,
  1579

\bibitem[{{Younger} {et~al.}(2008){Younger}, {Besla}, {Cox}, {Hernquist},
  {Robertson}, \& {Willman}}]{Younger2008}
{Younger}, J.~D., {Besla}, G., {Cox}, T.~J., {et~al.} 2008, \apjl, 676, L21

\end{thebibliography}

\begin{table}
\begin{center}
\caption{Summary of Observations}
\begin{tabular}{@{}lrrcrccc}
\hline \hline
Object & $\alpha_{0}$ (h~m~s) & $\delta_{0}$ (d~m~s) & Telescope & Mosaic & $<X_{g}>$   &  $<X_{r}>$ & Exp. Time: $g/r$ (s)\\
\hline
   Eridanus     	& $04:24:44.50$ & $-21:07:42.9$	&CFHT	& $1\times1$ 	&  $1.33$ & $1.33$ 	& $6\times 270/ 6\times 270$\\
  NGC\,2419  	& $07:38:08.50$ & $38:56:24.9$ 	& CFHT	& $1\times1$ 	&  $1.40$ & $1.17$	& $6\times 450/ 6\times 450$\\
  Koposov\,2     & $07:58:17.00$ & $26:18:48.0$ 	&CFHT	& $1\times1$ 	&  $1.28$ & $1.26$ 	& $6\times 500/ 6\times 500$\\
   NGC~7006     & $21:01:29.50$ & $16:14:45.1$ 	& CFHT	& $1\times1$ 	&  $1.01$ & $1.01$ 	& $6\times 240/ 6\times 240$\\
\hline
\label{t:list}
\end{tabular}
\end{center}
\end{table}

\clearpage

\end{document}